# Developer Belief vs. Reality:
# The Case of the Commit Size Distribution

Dirk Riehle, Carsten Kolassa, Michel A. Salim

Friedrich-Alexander-University Erlangen-Nürnberg
Martensstr. 3, 91058 Erlangen, Germany
dirk@riehle.org, carsten@kolassa.de, michel@sylvestre.me

**Abstract:** The design of software development tools follows from what the developers of such tools believe is true about software development. A key aspect of such beliefs is the size of code contributions (commits) to a software project. In this paper, we show that what tool developers think is true about the size of code contributions is different by more than an order of magnitude from reality. We present this reality, called the commit size distribution, for a large sample of open source and selected closed source projects. We suggest that these new empirical insights will help improve software development tools by aligning underlying design assumptions closer with reality.

## 1. Introduction

In this paper, we study what developers believe is true about the size of code contributions to software development projects and how it compares with reality. We find that 1-3 lines of code are the most common commit size while developers predict it would be much higher. The contributions of this paper are:

- The results of a developer survey on commit sizes and an assessment of the beliefs developers hold about them,
- the assessment of the reality of commit sizes in software development for a large sample of open source software projects,
- the assessment of the reality of commit sizes in software development for selected closed source software projects of SAP, a large software vendor,
- the surprising conclusion that software developer beliefs differ significantly from development reality.

Finding that there is a significant difference between reality and what developers believe about code contributions leads us to suggest that software development tools may have suboptimal designs that can be improved with our newly found knowledge. Our work strongly suggests revisiting the design of code-centric development tools.

Our finding may be surprising, depending on which school of psychology you belong to. By now classic research into cognitive biases [6] has shown that humans are poor estimators and various biases get in the way of accurately estimating properties of distributions like the one discussed in this paper [15]. However, recent research also has shown that under defined experimental situations, people can be capable of predicting the mode and percentiles of a distribution well [12] [4]. In the discussion section, we offer some possible explanations. However, it is not the purpose of this paper to explain this finding but only to bring it into the open so that we can build better software development tools.

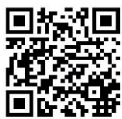

The paper is structured as follows. Section 2 defines the necessary terms. Section 3 reports the results of a developer survey to understand what developers believe is true about commit sizes in software development. Section 4 then shows the measurable reality for open source and a few selected closed source projects. Section 5 compares developer belief with reality and shows that beliefs differ from reality by at least an order of magnitude. Section 6 discusses threats to the validity of our analysis. Section 7 discusses related work and Section 8 concludes the paper.

## 2. Commit Sizes

Software developers contribute code to software projects when going about their programming work. Such code contribution is commonly called a "commit". Any such commit adds, removes, or changes existing source code. A common measure of size for commits is lines of code (LoC), approximating the amount of work spent on the commit. We distinguish source code lines from comment lines from empty lines:

- A *source code line* (SLoC) contains program code,
- a *comment line* (CL) contains only comments, and
- an *empty line* contains only whitespace.

We define LoC = SLoC + CL and thus measure the size of a commit in source code lines + comment lines. Measuring the size of a commit is non-trivial. The primary tool for assessing commit sizes is the "diff" tool, which compares source code files and tells its user which lines have been added and which lines have been removed.

- A *commit* is the sum of all diffs over all affected files where
- a *diff* lists several diff chunks in one file, where
- a *diff chunk* identifies co-located lines added or removed.

Unfortunately, a diff tool cannot identify with certainty that a line was changed; such a changed line is always counted as one line removed and one line added. However, a changed line should count as one line of work, while an added and a removed line of code should count as two lines of work. Enhanced diff algorithms exist, for example [3], that use text analysis to calculate the probability of whether one line added and one line removed equals one line changed or two separate lines, one added, one removed. However, these algorithms can't provide certainty and are computationally expensive.

Prior work determined that for large numbers of commits, the mean of the minimally and maximally possible values is a statistically valid estimate for the diff chunk size [9]. In this work, we calculate commit sizes for more than 8 million commits. Table 1 provides the resulting equations: "a" represents the number of lines added and "r" represents the number of lines removed according to the diff tool. We compute commit sizes by adding up the diff chunk sizes computed using equation (c) of Table 1.

*Table 1: Equations used to compute a commit's size*

(a)  lower_bound(a, r) = max(a, r)                    // full overlap, highest number of changed lines
(b)  upper_bound(a, r) = a + r                         // no overlap, no changed lines in diff chunk
(c)  diff_chunk_size(a, r) = (lower_bound(a, r) + upper_bound(a, r)) / 2      // mean of both bounds

The *commit size distribution* of some commit population is the distribution of the number of occurrences (y-axis) of all possible commit sizes (x-axis). The *commit size distribution of open source* has all commits in all open source projects as the population, and the *commit size distribution of closed source* has all commits of all closed source projects as the population.

## 3. Developer Belief

To assess what developers believe is true about commit sizes, we performed a survey in early 2010. We asked about the commit size distribution of all of open source and all of closed source, respectively. We asked what developers thought was

- the "most frequent commit size" (mode),
- the "average commit size" (mean),
- the commit size at the $90^{th}$ percentile, and
- the shape of the commit size distribution.

We asked these questions separately for all of open source and all of closed source software, and we assessed survey respondent demographics to the extent that it was relevant to evaluating this survey. An underlying assumption is that different projects have similar commit size distributions. We made this assumption explicit to survey respondents and Section 4 validates it. The survey is available on the web [14].

We performed pre-tests to ensure that definitions are clear and will be understood by survey respondents. We reached out to software developers using various channels, mostly mailing lists, but also social media tools like blogs, micro-blogging services, and social networking services. We received 73 valid and complete survey responses. In the threats to validity section we look at the sampling error and show why we believe that this low return rate does not affect the validity of our conclusions.

Figures 1-4 show developer beliefs for mode and $90^{th}$ percentile of the open source or closed source commit size distributions. The $90^{th}$ percentile commit size defines the size that respondents believe is larger than 90% of all other commit sizes. We omit the discussion of the mean in this Section, because the survey results are in line with those of the mode and and $90^{th}$ percentile and don't add anything to the discussion.

Figures 5-6 show what survey respondents believe is the difference between open source and closed source. Between 10%-20% of all respondents believe that there is no difference between open source and closed source as to mode and $90^{th}$ percentile. Thus, the vast majority believes there is a difference. In general, survey respondents believe that open source has a smaller mode and $90^{th}$ percentile of commit sizes than closed source.

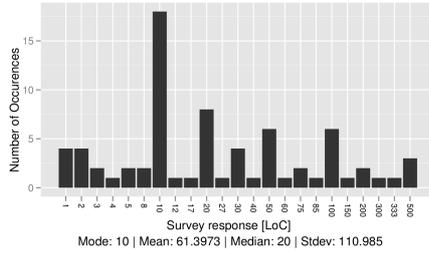

*Figure 1: Survey results of developer belief for mode of open source commit size distribution*

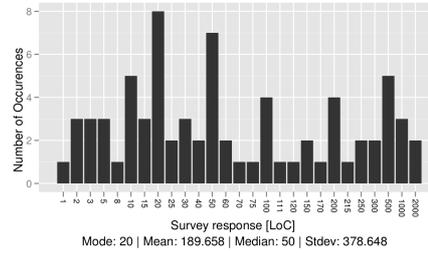

*Figure 2: Survey results of developer belief for mode of closed source commit size distribution*

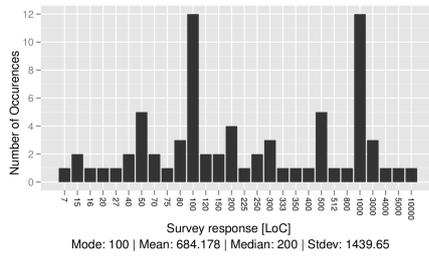

*Figure 3: Survey results of developer belief for 90[th] percentile of open source commit size distribution*

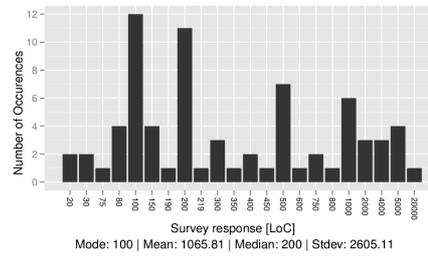

*Figure 4: Survey results of developer belief for 90[th] percentile of closed source commit size distribution*

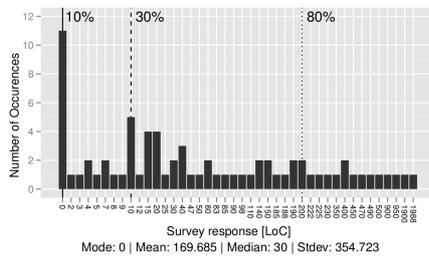

*Figure 5: Absolute value of difference in developer belief between mode of open and closed source commit size distribution*

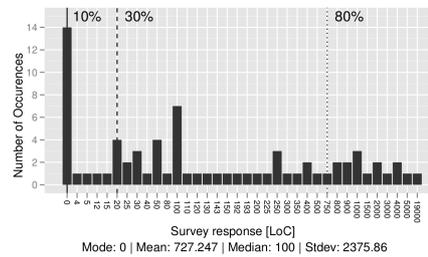

*Figure 6: Absolute value of difference in developer belief between 90[th] percentile of open and closed source commit size distribution*

We visualized three different distributions and let survey respondents choose which one they felt was closest to reality. The three alternatives were (a) a normal distribution, (b) a distribution following a power law, and (c) a distribution skewed towards large commits. Tables 2 and 3 show how respondents chose among the alternatives. There was no clear consensus and all three distributions had roughly equal support among respondents.

*Table 2: Survey results of developer belief for shape of open source commit size distribution*

| | |
|---|---|
| Normal distribution: | 16% |
| Power law distribution: | 59% |
| Skewed to large commits: | 25% |

*Table 3: Survey results of developer belief for shape of closed source commit size distribution*

| | |
|---|---|
| Normal distribution: | 35% |
| Power law distribution: | 33% |
| Skewed to large commits: | 32% |

## 4. Development Reality

Open source software development is usually public software development. Thus, for open source, we can know the commit size distribution and do not have to rely on beliefs. For closed source software development we have to rely on selected projects as the complete commit size distribution of closed source or even a representative sample is practically impossible to determine.

### 4.1 The Open Source Distribution

To assess the commit size distribution of open source, we use a database snapshot of the Ohloh.net open source project data website. Our snapshot is dated March 2008. Unlike data sources like SourceForge.net [10] the Ohloh data has no apparent bias towards any particular category of open source project. The only bias we could see is a focus on active projects with engaged user communities, as the Ohloh service requires community participation to have a project listed (self-reporting bias), as well as an English-language bias, given that the Ohloh website is written in English.

Our snapshot contains 11,143 open source projects. In September 2007, Daffara estimated that there are 18,000 active open source projects in the world [5]. (The total number of projects is much larger, but most open source projects are not active and by our activity definition have to be excluded.) Using the same definition of "active project" as Daffara our database snapshot contains 5,117 active open source projects. Thus, we estimate that our database contains about 30% of all open source projects active in March 2008.

The Ohloh database contains the complete commit history of all of its projects to the extent that it is available on the web, going back as far as 1991. Using the definitions from Section 2, we measure the commit size distribution of our open source sample population. Iterating over all 8,705,118 commits in the database, we compute the commit sizes and determine the commit size distribution.

Figure 7 shows the commit size distribution of our open source sample. The distribution is shown as a probability density function (PDF) for visualization and comprehension purposes. Integrating the interval [0, 1] provides the probability of a commit of size 0 or 1 LoC, [0, 2] provides the probability of a commit of size 0, 1, or 2 LoC, etc. The PDF is strictly falling with 1-3 LoC being most frequent.

The PDF presented here is empirical: It does not present an analytically closed model; all that is being presented is measured data. In other work we show that a generalized Pareto distribution fits the empirical data well, but this discussion is omitted here for reasons of space. As a Pareto distribution, the shape of the distribution follows a power law.

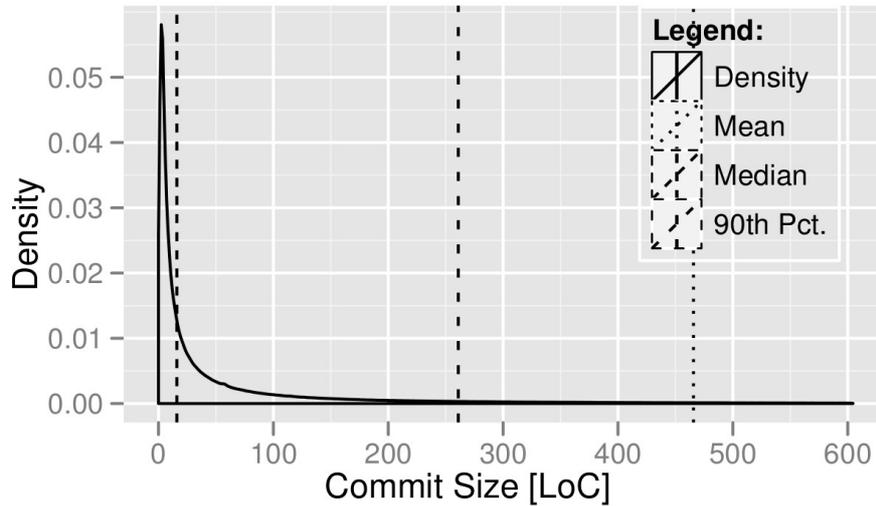

*Figure 7: The (empirical) PDF of the commit size distribution of about 30% of all active open source projects (March 2008). The graphs in Figures 7-9 have been calculated using R and ggplot2, which uses a Gaussian smoothing kernel with the standard deviation as bandwidth.*

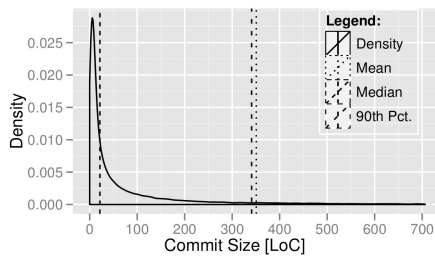

*Figure 8: The PDF of the commit size distribution of the closed source BAS project at age 10-19 years*

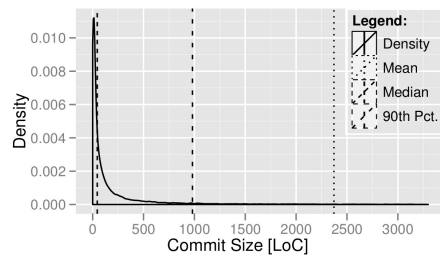

*Figure 9: The PDF of the commit size distribution of 122 closed source projects at age 0-3 years*

The mode of the distribution is 1.5 LoC. This real number is an artifact of measuring commit sizes using the estimation algorithm discussed in Section 2. An added and a removed line of code maps as often on one changed line of code as on one added and one separately removed line of code, averaging out to 1.5 LoC.

**4.2 Closed Source Distributions**

This section presents an analysis of selected projects of SAP AG, one of the world's largest software producers. We retrieved data for two types of projects:

- SAP's core virtual machine and libraries, the BAS project, at age 10-19 years, and
- 122 of SAP's research projects during their first three years of life.

We chose these projects due to the stark discrepancy between the two types of projects. The core virtual machine and libraries (the "BAS" project, short for base) is perhaps the most robust and thoroughly tested code SAP has ever developed and it forms the base of its software stack. It is written in C. The commit count of the BAS project is 56,840 commits and the configuration management system is Perforce.

The 122 research projects are young projects with a high mortality rate. The code base is undergoing wild changes and rapid development. They are written in C-style languages like C, C++, and Java. The commit count for the projects is 23,271 commits and the configuration management system is Subversion.

Figure 8 shows the PDF of the distribution of SAP's BAS project and Figure 9 shows the PDF of the distribution of the 122 research projects. The research projects all had similar distributions and like for open source have been grouped as one.

The commit size distribution of the BAS project is similar to the one of the research projects. This may come as a surprise given that a hypothesis might have been that mature projects develop more incrementally and using smaller code changes than young projects, which might move forward in (code) leaps and bounds. However, even young research projects apparently are moving forward mostly in small incremental commits.

Thus, the commit size distributions of (a) about 30% of open source (at its time), of (b) a mature closed source project, and of (c) 122 young research projects undergoing rapid change all have similarily-shaped strictly falling distributions that follow a power law, with 1-3 lines of code being most common. The actual parameters of the distributions differ, however, between open source and our closed source sample.

## 5. Belief vs. Reality

Table 4 summarizes the differences between reality and what developers said about commit size distributions. As to the mode of the distribution, the most frequent commit size, survey respondents are off by a wide margin. In the case of open source, the predicted value is 20 LoC with a measured value of 1.5 LoC and a relative error of 12. In the case of closed source, the predicted value is is 50 LoC with a real value of 3 LoC and a relative error of 16. Thus, respondents overestimate the mode significantly.

In general, the average developer believes that the commit size distribution has its maximum at around 20 LoC (open source) and 50 LoC (closed source) and then falls off rapidly as commit sizes increase. In reality, the commit size distributions of open source and closed source are strictly falling but have a longer tail than expected, with the mean value and $90^{th}$ percentile farther out than what developers are expecting.

The long tail of the distribution, both open source in general and selected closed source projects, becomes evident by the $90^{th}$ percentile of commit sizes being lower than the mean value of the commit sizes: This implies that a significant number of large commits are regularly being contributed, more than anticipated by survey respondents. The mode of the open and closed source distributions are close to each other, suggesting little practical difference between open and closed source projects. The mean and the 90[th] percentile are different between open and closed source. Also, survey respondents underestimated the mean significantly.

One might argue that it is hard even for technically schooled people to distinguish mode from mean from median. However, in our survey [14] we did not use these words, but visualized their meaning. We believe that respondents had a sufficiently good understanding of these terms when making their choices.

*Table 4: Belief, reality, absolute, and relative error for open and closed source distributions*

|  | 1. Open Source (Large Sample) | | | | 2. Closed Source (Selected Projects) | | | | 3. Difference Between Open & Closed Source | | | |
|---|---|---|---|---|---|---|---|---|---|---|---|---|
|  | Belief (median) | Reality | Absolute Error | Relative Error | Belief (median) | Reality | Absolute Error | Relative Error | Belief (median) | Reality | Absolute Error | Relative Error |
| **Mode [LoC]** | 20 | 1.5 | 18.5 | 12.33 | 50 | 3 | 47 | 15.67 | 30 | 1.5 | 28.5 | 19 |
| **Mean [LoC]** | 50 | 465.7 | 415.7 | 0.8926 | 80 | 942.0 | 862.0 | 0.9151 | 55 | 476.3 | 421.3 | 0.8845 |
| **90[th] Pctle [LoC]** | 200 | 261 | 61 | 0.2337 | 200 | 461.5 | 261.5 | 0.5666 | 100 | 200.5 | 100.5 | 0.5012 |

*Table 5: Sampling error in survey respondents' answers when viewed as a random sample of the total developer population*

|  | 1. Open Source (Large Sample) | | | | 2. Closed Source (Sel. Projects) | | | |
|---|---|---|---|---|---|---|---|---|
|  | 25[th] Percentile | | 50[th] Percentile | | 25[th] Percentile | | 50[th] Percentile | |
|  | Absolute Error | Relative Error | Absolute Error | Relative Error | Absolute Error | Relative Error | Absolute Error | Relative Error |
| **Mode [LoC]** | 8.5 | 5.667 | 18.5 | 12.33 | 12.00 | 4.000 | 47.00 | 15.67 |
| **Mean [LoC]** | 365.7 | 0.7853 | 415.7 | 0.8926 | 742.0 | 0.7877 | 864.5 | 0.9177 |
| **90[th] Percentile [LoC]** | 161.0 | 0.6169 | 211.0 | 0.8084 | 261.5 | 0.5666 | 361.5 | 0.7833 |
| **Sampling Error [%]** | 10.00 (9.333) | | 11.55 (11.47) | | 10.00 | | 11.55 | |

## 6. Threats to Validity

We address a possible sampling error of survey responses, age of open and closed source data, issues with measuring commit sizes and representativeness of project data.

### 6.1 Survey Responses

With 73 valid and complete survey responses, the response rate is on the low side. However, from the demographics gathered in the survey, there was no apparent bias. Thus, the question of the representativeness of the survey becomes one of sampling error.

Table 5 above presents the difference between the medians of the predicted and the real commit sizes. To show that the results of Table 5 are representative we first calculate the difference between the predicted and the real value for each survey respondent. Table 5 shows the 25th percentile and the 50th percentile of that difference.

For the 50th percentile, the sampling error for the survey responses is 11.55% at a confidence level of 95%. Thus, even in the worst case scenario more than 38% of all developers have an error above the 50th percentile. We also calculate the sampling error for the 25th percentile which is 10.00% at a confidence level of 95%. Thus, at least 65% of all developers are off by this error. This shows that even using conservative estimates we can consider the survey responses as representative.

**6.2 Age of Analysis Data**

Our Ohloh database snapshot is dated March 2008. However, for this analysis the age doesn't matter much. We have no indication that open source changed significantly from 2008 to today (2011). Hence we believe that adding three years to an analysis history of 15+ years will not change the results in any significant way.

**6.3 Measuring Commit Sizes**

We spent considerable effort on developing a well-performing algorithm for determining commit sizes (Section 2). The core equation can be computed fast but only provides an estimate of the size of a given commit. Thus, in any given instance it may be off, but when measured over large commit size populations, it will be accurate [9]. This is the situation we were facing in this paper. The commit size population is greater than 8 million commits and is sufficiently large to allow for the probabilistic estimation of commit sizes by the simple heuristic of Section 2 that we use throughout this paper.

**6.4 Representativeness of Project Data**

Our open source sample population is close to being representative of open source. With about 30% of all active open source projects at the time of the database snapshot, we have captured a significant chunk of the total population. As discussed, except for self-reporting and English-language bias, we find no apparent bias in our sample. Moreover, there is no apparent connection between this potential bias and the measure of interest, commit sizes. Thus, we believe that the commit size distribution of open source is close or identical to the one presented in this paper.

Our closed source sample population is from a single vendor only and hence not representative. The question becomes how far off from representative values are our closed source measurements? First, there is no significant difference between a mature and many young projects. Moreover, there is no significant difference between the general open source distribution and the commit size distribution of the closed source project sample. All of this is contrary to developer belief and our own prior expectations.

While we cannot overcome the fundamental problem the surprising similarity between the open source and closed source distributions suggests that to the extent that they are similar, the degree of representativeness of our open source sample applies to the representativeness of our closed source sample.

## 7. Related Work

We previously presented an analysis of the open source commit size distribution [2]. In comparison to that analysis, the work presented in this paper adds the survey, closed source data, and a more thorough investigation of the open source commit size distribution including using the more precise heuristic for estimating commit sizes of Section 2.

Alali et al. present an analysis of "a typical commit" using the version history of 9 open source projects [1]. They mostly focus on the number of files changed (and how), but also provide chunk and line-size data. They compute line size changes by adding lines added and removed, thus overestimating sizes by ignoring changed lines of code. Still, they find "quite small" commit sizes without giving more details. Interestingly, they find a strong correlation between diff chunk and size. Alali et al.'s 9 projects are large well-known open source projects. In contrast to Alali we focus solely on commit sizes, use a more precise measure and compute a derived function, the commit size distribution, on a close-to-representative sample rather than 9 selected projects.

Hattori and Lanza discuss "the nature of commits" in which they look at 9 different open source projects [7]. They measure commit sizes by number of files changed and find a distribution similar to ours. Beyond the distribution, their work is mostly about classifying commits while we focus on the difference between developer belief and reality. Similarly, Hindle et al. analyze 2000 large commits from 9 selected open source projects and find that small commits are more corrective while large commits are more perfective [8].

Purushothaman and Perry analyze the impact that small changes have on quality attributes of software under consideration [13]. Their data is derived from a single large closed source project. They find that one-line changes represent the majority of changes during maintenance, which is in line with our results. Implicit in the choice of research topics by Purushothama and Perry as well as Hindle et al. may be an assumption that commit sizes are smaller in maintenance than in development mode. Our closed source data does not immediately support such a hypothesis but it is worth investigating further.

Weißgerber et al. look at the patch submission and acceptance process of two open source projects [16]. They find that small patches are more likely to get accepted into the code base than large patches. An obvious reason may be that small patches are easier to review than large patches which, if not handled quickly, get harder to review and accept with time. While not representative, Weißgerber's observation is interesting to us, as it might explain why the commit size distribution of open source is falling more quickly than those of the closed source projects we analyzed.

## 8. Conclusions

The paper shows that the commit size distribution of open source runs counter to the expectations of respondents from a software developer survey we undertook. The only exception was the 90[th] percentile, which respondents were able to predict reasonably well. Survey respondents could not agree on the form of the distribution and expected a much higher mode for commit sizes in open source as well as closed source software development. They also expected the distribution to fall of more quickly than it does. Thus, developers underestimate the significance of small commits and don't realize the long tail of commit sizes. Moreover, respondents expected to see significant differences between open and closed source software development, and our data shows these differences aren't there.

The survey findings are particularly significant, because we found no difference of opinion between tool developers and regular developers. Thus, a reasonable assumption is that the conceptual model of software tool developers may not be well aligned with reality when it comes to the commit size distribution. Our work provides that reality and can be used to improve the design of software development tools. We can now rethink the use of screen real estate in merge tools, or the design and implementation of configuration management systems, or how to improve change impact analysis using the commit size statistics that our paper presents.

It is left for us to speculate why survey respondents were off to the observable extent. One possible answer is that commit sizes may have become smaller over time so that past experiences unduly affected respondents' judgment. Such cognitive bias is supported by psychology research that reports that humans hesitate to go to the extremes but rather remain conservative estimators [11]. Other explanations are possible as well, but we leave them to psychologists and future work.

## Acknowledgements

We would like to thank York Thomas of HPI for providing us with the SAP data. We would also like to thank Manuel Klimek of Google for helping us with the survey distribution. We would like to thank Wolfgang Mauerer of Siemens and Lutz Prechelt of FU Berlin for feedback on the paper. We would like to thank Oscar Nierstrasz and Niko Schwarz for helping us improve the paper and for contributing the references about psychology research on cognitive biases. We would like to thank all other reviewers for their help as well.